\documentclass{article}
\usepackage {graphicx,latexsym}
\usepackage{amsmath,amsthm}
\usepackage{amsfonts}
\usepackage[utf8]{inputenc}

\usepackage{caption}
\usepackage{subcaption}

\newcommand{\R}{{\mathbb R}}

\newtheorem{definition}{Definition}

\begin{document}

\title{
Observable currents\\ 
in lattice field theories
}
\author{
José A. Zapata
\footnote{e-mail: \ttfamily zapata@matmor.unam.mx} 
\\
{\it Centro de Ciencias Matemáticas,}\\ 
{\it Universidad Nacional Autónoma de México} \\ 
{\it C.P. 58089, Morelia, Michoacán, México}\\
}

\date{}
\maketitle

\begin{abstract} 
Observable currents are spacetime local objects that 
induce physical observables when integrated on an auxiliary codimension one surface. 
Since the resulting observables are independent of local deformations of the integration surface, 
the currents themselves carry most of the information about the induced physical observables. 
I introduce observable currents in a multisymplectic framework for Lagrangian field theory over discrete spacetime. 
One family of examples is composed by Noether currents. 
A much larger family of examples is composed by currents, spacetime local objects, 
that encode the symplectic product between two arbitrary vectors tangent to the space of solutions. 
A weak version of observable currents, which in general are nonlocal, 
is also introduced. Weak observable currents can be used to 
separate points in the space of physically distinct solutions. 
It is shown that 
a large class of weak observable currents can be ``improved'' to become local. 
A Poisson bracket gives the space of observable currents the structure of a Lie algebra. 
Peierls bracket for bulk observables gives 
an algebra homomorphism mapping equivalence classes of 
bulk observables to weak observable currents. 
The study covers scalar fields, nonlinear sigma models and gauge theories (including gauge theory formulations of general relativity) on the lattice. 
Even when this paper is entirely classical, this study is relevant for quantum field theory because 
a quantization of the framework leads to a spin foam model formulation of lattice field theory.

\end{abstract}

%
%
%
%
%
%
%
%
%
%
%


\section{Objective and results}

Physical observables in covariant field theory are usually constructed as an 
integral of a function of the field and its derivatives weighted with a smearing function 
and evaluated on a solution of the field equations.  
This procedure of generating physical observables produces an equivalence relation among functions on the space of histories; two functions are declared equivalent if they agree when restricted to the space of solutions. 
Beyond this, zeroth-order reason to declare an equivalence relation there is a first-order reason: 
one may want to identify bulk observables inducing the same vector field on the space of solutions via Peierls bracket. 
In contrast, no equivalence relations are needed 
if one does not work in a covariant formalism, but instead one works in a space of initial data given by a choice of a Cauchy surface (for globally hyperbolic theories). A physical observable may be defined as a function in the space of solutions induced by a function in the space of initial data. 
Moreover, 
the Poisson bracket of functions of initial data turns out to be again a function of initial data. 
On the other hand, working with initial data to encode solutions means that one is tied to solving the field equations at least implicitly, 
and it also means surrendering the access to a natural notion of spacetime locality by trading it for a notion of locality inherited from the Cauchy surface. Because of these considerations, 
I use a spacetime covariant framework but study physical observables generated from integration on codimension one surfaces. 
Since locality is a primary motivation of this study, I work within the multisymplectic formulation of Lagrangian field theory where the primary ingredients are local objects in a finite dimensional manifold \cite{MultisympCont}. 
This framework takes place in the manifold that encodes spacetime position, the field's value and the field's partial derivatives -- the first jet bundle $J^1 Y$.

I will define observable currents within a version for multisymplectic field theory over a discrete spacetime \cite{meffGBFT}. 
Even when this paper is entirely classical, this study is relevant for quantum field theory because 
a quantization of the mentioned framework leads to a spin foam model formulation of lattice field theory.
After integration (really a summation in the discrete scenario) 
on a codimension one surface, they will yield physical observables. 
The resulting expression will be a function that could be evaluated on any history (local section), but if the history happens to be a solution of the field equations, the integration surface will become a mere auxiliary object in the sense that the outcome of evaluation will be invariant under local deformations of the surface. A precise statement is that the integration surface will only be relevant up to homology. 
Within the multisymplectic ideology, this is the natural structure to consider. Let me explain. 
The cornerstone of multisymplectic geometry is a generalization the conservation of the symplectic form that is appropriate for covariant field theory. The structural conservation law is called the multisymplectic formula, and one of its consequences is that to any oriented codimension one surface $\Sigma$, the the multisymplectic form $\Omega_L \in \Lambda^{n+1}(J^1 Y)$ assigns a presymplectic form in the space of solutions 
$\bar{\omega}_\Sigma \in \Lambda^2({\rm Sols}_U)$
that is largely independent of the integration surface (it depends only on $\Sigma$'s homology class). 
This looks good, but what is remarkable about this approach to field theory is that in the formulation of  
the conservation law of the presymplectic structure, 
spacetime locality can be kept at the forefront instead of formulating it in the space of solutions 
as stated in the previous sentence. The assignment of presymplectic forms to codimension one surfaces is compatible with the solutions to the field equations; this compatibility generalizes the conservation law of the presymplectic structure and is captured in 
a local equation in $J^1 Y$ 
-- the multisymplectic formula. 
Thus, when considering the equation 
$df = - \iota_{\bar{v}}\bar{\omega}_\Sigma$
for the Hamiltonian vector field $\bar{v} \in {\mathfrak X}({\rm Sols}_U)$
associated with a function of solutions $f\in {\cal C}^\infty ({\rm Sols}_U)$ and noticing 
that the right-hand side is calculated from local objects in the jet, it 
is only natural to look for a spacetime local relation that originates it. 
I will consider 
functions that are also calculated from local objects in the jet, 
$f_\Sigma(\phi) = \int_\Sigma (j^1\phi)^\ast F$, 
where $F$ is an $n-1$ form on the jet satisfying conditions that 
let it participate in the local equation 
\[
dF = - \iota_{\tilde{v}} \Omega_L
\]
matching the conservation properties of the right hand side. 
That is, when evaluated on solutions, $f_\Sigma(\phi)$ is independent of local deformations of 
$\Sigma$ in the sense that only its homology type is relevant.  
This paper's objective is to explore this theme in a discrete field theory framework. 

The main results of the paper are outlined below: 
\begin{itemize}
\item
Observable currents are {\em spacetime local} objects that 
induce functions of histories after they are integrated on codimension one surfaces. 
When evaluated on solutions the integration surface may be locally deformed by without modifying the outcome. 
One family of examples is composed by Noether currents. 
A much larger family of examples is composed by currents, spacetime local objects, 
that encode the symplectic product between two arbitrary vectors tangent to the space of solutions. 
\item
If the notion of observable current is appropriately weakened, every physically relevant local property of the field is measurable by the resulting physical observables. 
I will prove, within a framework for lattice field theories, 
that {\em this family of physical observables 
distinguishes physically distinct solutions of the field equations.} 
This might not be surprising since weak observable currents are not local objects that are defined only on solutions. 
However, the strength of the result and the relation between weak observable currents and observable currents will make the result useful. 

The differential of a weak observable current 
can be calculated from a restriction of 
the same local object that may be the differential of an observable current. 
A important class of weak observable currents can be promoted to observable currents: the elements of the class are those with an associated Hamiltonian vector field that is a commutator (or linear combination of commutators). 
For some field theories the mentioned class may be so large that 
observable currents are enough to distinguish physically distinct solutions. 
\item
There is a class of observable currents that participate in the local equation that in the space of solutions induces the equation $df = - \iota_v \omega$, in which the (pre)symplectic form links functions on the space of solutions to vector fields. The local (multisymplectic) equation is $dF=-\iota_{\tilde{v}} \Omega_L + \alpha_{\tiny \Sigma}$, 
where the term $\alpha_{\tiny \Sigma}$ is innocuous in 
the sense that it does not affect the mentioned induced relation in the space of solutions, 
but its presence allows for a large class of Hamiltonian 
observable currents that would not be so without it. 
This new version of the equation linking observable currents with vector fields 
can also be used in the continuum \cite{OCsCont}. 
\item
A Poisson bracket for observable currents follows naturally from the multisymplectic structure making the space of observable currents a Lie algebra. 
Moreover, a product among observable currents together with the bracket make the structure a Poisson algebra. 
Consequently, a covariant Poisson structure for physical observables in lattice field theory has been introduced. 
\item
The bracket of Peierls \cite{Peierls} for bulk observables (adapted to the lattice) 
would induce an algebra homomorphism mapping equivalence classes of 
bulk observables to weak observable currents. 
Thus, one may say that 
{\em the algebra of weak observable currents captures the essence of the algebra of physical observables of the theory}. 
\end{itemize}

The Poisson structure arising from the continuum time Hamiltonian approach of Kogut and Susskind 
\cite{Kogut-Susskind} should be related to that presented here, after a continuum limit in time and a Legendre transformation are performed. This subject merits closer study. 

In the multisymplectic framework for classical field theories, there has been much debate regarding physical observables, and there are several proposals for them (see\cite{MSobs} and references therein). In particular, observables arising from the integration of $n-1$ forms have been considered. 
The definition of observable currents given in this paper has a similar spirit to that of  
``observable $n-1$ forms,'' and 
``dynamical observables'' appearing in the work of Hélein and Kouneiher and previously in the work of Kijowski, Szczyrba and Tulczyjew \cite{DynObsEtc}. Also the Hamiltonian observable currents introduced here  
are closely related to the ``algebraic observables'' considered in the same references. 
A more general proposal by Kanatchikov extends the study of ``algebraic observables'' to forms of arbitrary degree \cite{Kanatchikov}. 
However, the mentioned studies of observables within the multisymplectic framework for classical field theories in the continuum do not 
contain notions related to the weak observable currents introduced here, nor do they contain the extended notion of Hamiltonian observable currents 
introduced in this article. The mentioned new notions are directly responsible for reaching the strong results claimed above. 
A study of observable currents within the multisymplectic framework in the continuum is in preparation \cite{OCsCont}.

In Section \ref{Review}, I review the geometric framework for first-order Lagrangian classical 
lattice field theories that is used in the rest of the article. 
Section \ref{OCsSection} contains the main definitions and results. 
Section \ref{Bulk+Peierls} talks about the relation to the Peierls algebra of physical observables modeling 
bulk measurements. 
Section \ref{CoarseGraining} deals with the issue of coarse graining observable currents. 
I conclude with a short section including a summary of the results 
and final remarks.

\section{Review of a geometric framework for \\
classical lattice field theories}
\label{Review}
The main references leading to the formalism described below are: Veselov's Hamiltonian mechanics with discrete time \cite{Veselov}, a generalization of this work to field theory on a discrete spacetime due to Marsden et al \cite{MarsdenEtAl}, and Reisenberger's discretization of spacetime used in his first works on spin foam models and his lattice model for general relativity \cite{Reisenberger}. 
The formalism shares goals and ingredients with \cite{related}.

In this section, I give a short review of a geometric framework for first-order Lagrangian classical 
lattice field theories in the case of scalar fields taking values on a typical fiber ${\cal F}$ that may be a vector space or a nonlinear fiber. The formalism exists also for gauge theories and extended gauge theories (including general relativity). 
For a more detailed exposition of the framework for all types of fields, see 
\cite{meffGBFT}.

A first-order Lagrangian density is an $n$-form in spacetime with domain 
\[
{\cal L} = {\cal L} (x, \phi, D\phi ) .
\]
The first step then in developing a discrete framework for field theory is to give a discrete counterpart of the first jet bundle 
$J^1Y \ni (x, \phi, D\phi )$. Once we have it, we will be able to work in a finite dimensional manifold using local objects (differential forms). Histories will be local sections; among them we will have solutions to field equations. The geometric structure behind this framework for classical field theory will emerge as relation among the local objects that hold when evaluated on solutions. The most important structure is called the multisymplectic formula, which is a covariant generalization of the conservation of the symplectic structure of classical mechanics under the evolution dictated by the field equations.


With the aide of a triangulation $\Delta$ of spacetime, 
the compact connected region of our interest 
$U\subset M$ is divided into a collection of closed $n$ dimensional simplices which 
share $n-1$ dimensional faces with their neighbors. 
A generic $n$-simplex in $U$ is denoted by 
$\nu \in U_\Delta^n$, and a generic $n-1$ simplex in $U$ is denoted by $\tau \in U_\Delta^{n-1}$. 
The boundary of our region is assumed to be a $n-1$ manifold (possibly with corners) which may have several connected components. It inherits a triangulation, and its $n-1$ simplices will be denoted by 
$\tau \in (\partial U)_\Delta^{n-1}$. 

Our discretization of a first-order Lagrangian theory, in the case of scalar fields, is based on decimating each $\nu \in U_\Delta^n$ by keeping track of only one point in its interior 
$C\nu\in \nu^\circ$ ``representing the bulk of atom $\nu$,'' and also keeping track of one point 
$C\tau\in \tau^\circ$ per boundary face ``representing boundary face
$\tau \subset \partial \nu$.'' 
Our decimated description of the field records its value in the discrete set of points described above and organizes it as a local section $\phi$ of a fiber bundle over $\Delta$. Chains of dimension $n$ and $n-1$ in $\Delta$ are lifted by $\phi$ to chains in the bundle 
$U_\Delta^n \ni \nu \overset{\phi}{\longmapsto} (\nu , \phi_\nu) \subset Y_\Delta$, 
$U_\Delta^{n-1} \ni \tau \overset{\phi}{\longmapsto} (\tau , \phi_\tau) \subset Y_\Delta$. 
Similarly, local 1st order data is visualized as a local section $\tilde{\phi}$ lifting chains from $\Delta$ to a discrete version of the first jet bundle, and it is presented in a format that displays spacetime position, value of the field and first-order changes of the field estimated using values of the field over neighboring points
\begin{eqnarray*}
U_\Delta^n \ni 
\nu 
& \overset{\tilde{\phi}}{\longmapsto} &
(\nu, \phi_\nu , 
\{ ( \phi_\nu, \phi_\tau) \}_{\tau \subset \partial \nu}) 
\subset J^1 Y_\Delta
\quad , \\
U_\Delta^{n-1} \ni 
\tau_\nu 
& \overset{\tilde{\phi}}{\longmapsto} &
(\tau, \phi_\tau , 
( \phi_\tau ,\phi_\nu )) 
\subset J^1 Y_\Delta
\quad . 
\end{eqnarray*}
Notice that the 1st order data over a codimension one simplex $\tau$ used here, $\tilde{\phi}(\tau_\nu)$, is composed by the value of the field and a ``partial derivative'' normal to $\tau$ which depends also on a $n$ simplex $\nu$ containing $\tau$ as a face. 

\begin{figure}[h!]
  \centering
{%
      \includegraphics[width=0.8\textwidth]{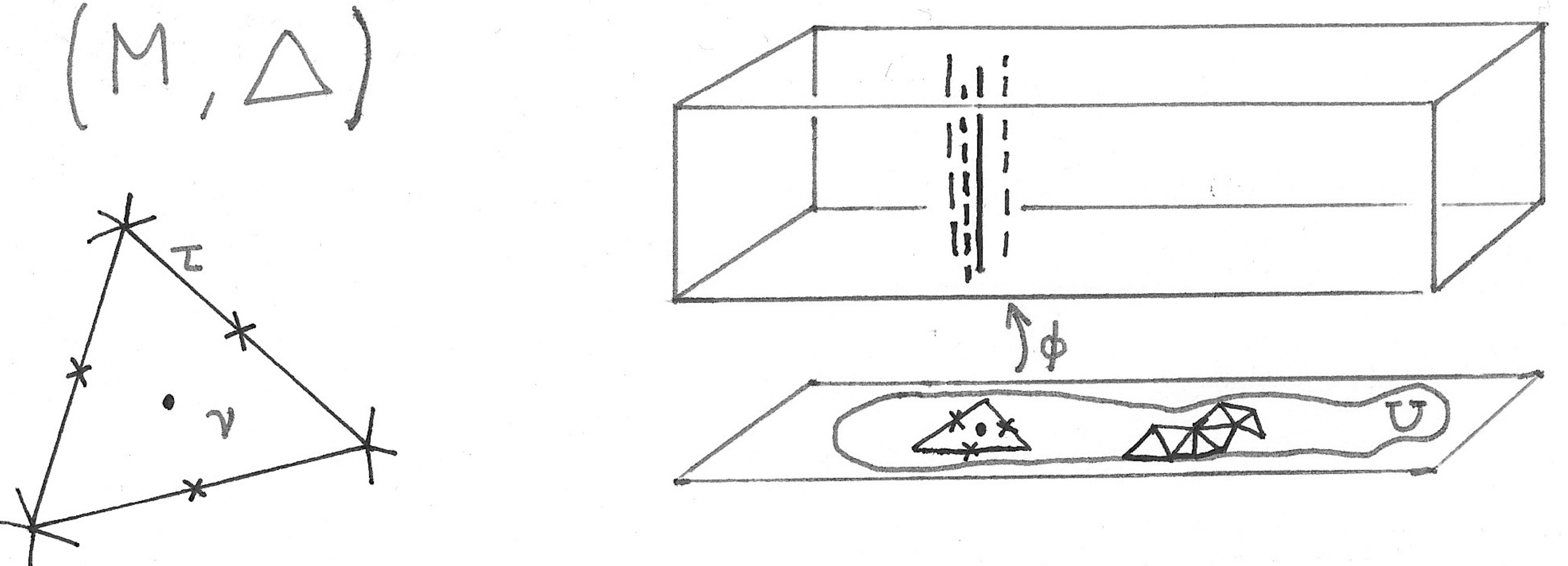}}
\end{figure}
We will work with vertical variations of the field $\phi \in {\rm Hists}_U$ that in our 1st order format are vector fields $\tilde{v}\in{\mathfrak X}(J^1 Y_\Delta |_U)$ displayed as 
\[
\delta\tilde{\phi}(\nu)=\tilde{v}(\nu)= (v_\nu \in T_{\phi_\nu}{\cal F}, 
\{ 
(v_\nu , v_\tau \in T_{\phi_\tau}{\cal F} )
\}_{\tau \subset \partial \nu}), 
\]
\[
\tilde{v}(\tau_\nu)= (v_\tau \in T_{\phi_\tau}{\cal F}, 
(v_\tau , v_\nu \in T_{\phi_\nu}{\cal F} ) ). 
\]

The action that will dictate dynamics has the form 
\[
S_U(\phi) = \sum_{\nu \in U_\Delta^n} 
L(\tilde{\phi}(\nu)) ,  
\]
where the discrete Lagrangian $L$ is thought of as a $n$ cochain in $J^1 Y_\Delta$

A variation of the history has an effect on the action 
that is calculated using finite dimensional calculus.  
The directional derivative of the action is simply a sum of partial derivatives of the Lagrangian: 
\[
dS_U(\phi) [v] = 
 \sum_U
\tilde{\phi}^\ast \; \iota_{\tilde{v}} ( {\rm E}_L + d_h \Theta_L ) , 
\]
where the sum is over all $n$ dimensional atoms $\nu \subset U$ and 
\[
\tilde{\phi}^\ast \; \iota_{\tilde{v}} {\rm E}_L = 
{\rm E}_L(\tilde{v} , \tilde{\phi}(\nu)) \doteq 
 \frac{\partial L}{\partial \phi}(\tilde{\phi}(\nu)) d\phi_\nu [\tilde{v}], 
\] 
\[
d_h \Theta_L (\tilde{v} , \tilde{\phi}(\nu)) \doteq 
\Theta_L (\tilde{v} , \tilde{\phi}(\partial \nu)) \doteq 
\sum_{\tau \in (\partial \nu)^{n-1}} \Theta_L (\tilde{v} , \tilde{\phi}(\tau_\nu)) , 
\]
\[
\Theta_L(\tilde{v} , \tilde{\phi}(\tau_\nu)) \doteq 
\frac{\partial L}{\partial \phi_\tau}(\tilde{\phi}(\nu)) d\phi_\tau [\tilde{v}]. 
\]
The first entry of this discrete version of Cartan's form $\Theta_L(\cdot, \cdot)$ is a one-form on the direction 
of the fibers, and its second entry has the interpretation of being a $n-1$ cochain on $J^1Y_\Delta$ because $\tau$ can be seen as a $n-1$ chain in $U$ and the section $\tilde{\phi}$ takes it to the discrete jet bundle. 
Note that in the second entry of $\Theta_L$ we have 
$\tilde{\phi}(\tau_\nu)$ where the subindex $\nu$ indicates that the partials of the Lagrangian are to be evaluated at $\tilde{\phi}(\nu)$ and not at $\tilde{\phi}(\nu')$ for $\nu'$ the other spacetime atom containing $\tau$. 
In this sense, the discrete version of Cartan's form is not strictly speaking a cochain.

Notice that a partial derivative of the Lagrangian with respect to 
variables of the type $\phi_\nu$, associated with the value of the field at the interior of the atoms, appears in only one term of $dS_U$. In contrast, 
a partial derivative of the Lagrangian with respect to 
$\phi_\tau$ appears in one term 
of $dS_U$ if $\tau \subset \partial U$ and appears in two terms if $\tau = \nu \cap \nu'$ 
is in the interior of $U$. 
Thus, Hamilton's principle yields field equations of the two types shown below: \\
Field equations internal to each $\nu$, 
asking the field over $\nu$ to be an extremal of the action, 
\begin{equation}
\label{internalFEQ}
{\rm E}_L(\cdot , \tilde{\phi}(\nu)) = 0 .
\end{equation}
Gluing (momentum matching) field equations at each $\tau = \nu \cap \nu'$ interior to $U$, 
asking that the solutions over $\nu$ and $\nu'$ have compatible momentum flux through 
$\tau_\nu = -\tau_{\nu'}$ (taking the orientation into account),  
\begin{equation}
\label{gluingFEQ}
\Theta_L(\cdot , \tilde{\phi}(\tau_\nu)) + \Theta_L(\cdot , \tilde{\phi}(\tau_{\nu'})) = 0.
\end{equation}

These field equations are algebraic conditions which select the solution submanifold of the space of histories ${\rm Sols}_U \subset {\rm Hists}_U$. In this formalism, on a compact portion of a discretization of spacetime, both of these spaces have a large, but finite, dimension. 
The space of first variations of a solution $\phi \in {\rm Sols}_U$ is the vector space $T_\phi {\rm Sols}_U \subset T_\phi {\rm Hists}_U$. Its elements $v \in T_\phi {\rm Sols}_U$ may be induced by vector fields along the fibers of $Y_\Delta$ that, when displayed in first-order format, become vector fields on $J^1Y_\Delta$ (in a vicinity of $\tilde{\phi}(U)$) satisfying the linearized field equations 
\begin{equation}
\label{linearizedFE}
{\cal L}_{\tilde{v}} E_L(\cdot , \tilde{\phi}(\nu))=0  , \quad
{\cal L}_{\tilde{v}} ( \Theta_L(\cdot , \tilde{\phi}(\tau_\nu)) + \Theta_L(\cdot , \tilde{\phi}(\tau_{\nu'})) )=0. 
\end{equation}
Vector fields in $J^1 Y_\Delta |_U$ 
satisfying these equations will also be called first variations, and the space of such vector fields will be denoted by ${\mathfrak F}_{\tilde{\phi}(U)} \subset {\mathfrak X}(J^1 Y_\Delta |_U)$. 

It will also be useful to give a bit of structure to the space of histories off-shell. 
I will assume that in the typical fiber ${\cal F}$ a Lie group $G$ has a transitive and free action. 
For example, if the fiber is $\R^n$, the action of $\R^n$ on itself by translations is the action that I am referring to. 
The field equations induce a stratification in the space of histories, 
${\rm Hists}_U = \cup_J {\rm Hists}_U^J$, 
\begin{equation}
\label{Jfieldeqs}
{\rm E}_L(\cdot , \tilde{\phi}(\nu)) = J_0(\cdot , \nu) , \quad 
\Theta_L(\cdot , \tilde{\phi}(\tau_\nu)) + \Theta_L(\cdot , \tilde{\phi}(\tau_{\nu'})) = J_1(\cdot , \tau),
\end{equation}
where the differential one-form valued $n$ cochain $J_0$ measures how much the internal field equations are broken, and 
the differential one-form valued $n-1$ cochain $J_1$ measures how much the gluing field equations are broken. 
The notation $J_0(\cdot , \nu)$ (respectively $J_1(\cdot , \tau)$) 
is meant to emphasize that by definition $J$ is assumed to be field independent in the sense that when acting on a $G$ invariant vector field the result is constant in the fiber ${\cal F}_\nu$ (respectively ${\cal F}_\tau$) 
and that the mentioned constant is independent of the field value in other fibers too. 
In the case of a gauge theory, the conventions are such that the equations written above are gauge covariant. 
The tangent spaces to each stratum $T_\phi {\rm Hists}_U^J$ is generated by vector fields satisfying the appropriate linearized field equations 
${\mathfrak F}_{\tilde{\phi}(U)}^J \subset {\mathfrak X}(J^1 Y_\Delta |_U)$. 

Hamilton's principle also leads us to geometric structure. 
The (pre)multisym\-plectic form 
\[
\Omega_L ( \tilde{v}(\nu), \tilde{w}(\nu) , \tilde{\phi}(\tau_\nu)) \doteq - d(\Theta_L|_{\tilde{\phi}(\tau_\nu)}) (\tilde{v}(\nu), \tilde{w}(\nu) ) 
\]
assigns (pre)symplectic structures 
(closed two-forms, but possibly degenerate) to spaces of data over 
oriented codimension one domains $\Sigma \mapsto \Omega_\Sigma$ 
\[
\omega_{L, \Sigma} 
(\tilde{v} , \tilde{w} ) = 
\sum_\Sigma \tilde{\phi}^\ast \; \iota_{\tilde{w}} \iota_{\tilde{v}} \Omega_L . 
\]
Notice here that the evaluation of $\Omega_L$ needs the specification of a codimension one simplex $\tau \in U_\Delta^{n-1}$ {\em and} a choice of an $n$ dimensional simplex $\nu \in U_\Delta^n$ that contains it. Every codimension one simplex in the interior of $U$ belongs to two ``atoms'' --it has two sides--; the orientation in $\Sigma$ is used to keep track of signs {\em and} to select a side. The convention is that if the orientation selects a normal vector to the surface, the side selected is in the oposite direction of the selected normal; in this way, the orientation of $\partial U$ induced by the orientation of $U$ (using outer normal convention) leads to a definition of $\omega_{L, \partial U}$ that is a sum partial derivatives of the Lagrangian evaluated in atoms interior to $U$.

Consider a solution of the field equations $\phi \in {\rm Sols}_U \subset {\rm Hists}_U$; evaluating $dS_U$ on it we obtain $dS_U(\phi) =  \sum_{\partial U} \tilde{\phi}^\ast \;  \Theta_L$. Since $d^2 =0$, for any vector fields corresponding to first variations 
$\tilde{v} , \tilde{w} \in {\mathfrak F}_{\tilde{\phi}(U)}$ the multisymplectic formula holds: 
\begin{equation}\label{MSformula}
\omega_{L, \partial U} (\tilde{v} , \tilde{w} ) = 
\sum_{\partial U} \tilde{\phi}^\ast \; \iota_{\tilde{w}} \iota_{\tilde{v}}
\Omega_L =0 . 
\end{equation}
Notice that the above conservation law holds for any $U' \subset U$; this follows from the same proof in which the action is restricted to $U'$. There is also a ``thin version'' of the multisymplectic formula that holds for any oriented codimension one surface $\Sigma \subset U$. For any solution $\phi \in {\rm Sols}_U$ and any pair of first variations 
$v, w \in {\mathfrak F}_{\tilde{\phi}(U)}$ the gluing field equations (\ref{gluingFEQ}) imply 
\begin{equation}\label{thinMSformula}
\omega_{L, \Sigma} (\tilde{v} , \tilde{w} ) + 
\omega_{L, \bar{\Sigma}}(\tilde{v} , \tilde{w}) 
= 0 , 	
\end{equation}
where $\bar{\Sigma}$ denotes the same surface with the opposite orientation. Notice that atoms in both sides of $\Sigma$ are involved in this equation. Thanks to the above equation, the validity of (\ref{MSformula}) in subregions of $U$ implies its validity in the resulting glued regions.

Consider an oriented codimension one surface $\Sigma$; the multisymplectic formula (\ref{MSformula}) says that there is an induced two-form in the space of solutions that is independent of local deformations of $\Sigma$ (more precisely, depending only on its homology class), $\bar{\omega}_{L, \Sigma} \in \Lambda^2({\rm Sols}_U)$. 

The multisymplectic formulas written above 
also hold off-shell when the vector fields $\tilde{v} , \tilde{w}$ belong to ${\mathfrak F}_{\tilde{\phi}(U)}^J$. 
If the fiber is non linear, even with the definition of a ``field independent'' $J$, 
the multisymplectic form needs to be modified off-shell to 
$\Omega_L + d J_1$. 
In order to prove the off-shell multisymplectic formulas, one can follow the same argument given above, but using an appropriately modified action. 
The reason is that given a history $\phi \in {\rm Hists}_U^J \subset {\rm Hists}_U$, 
the space ${\rm Hists}_U^J$ is the space of solutions of the modified action 
$S_U^J(\phi) = \sum_U (L(\tilde{\phi}(\nu)) - 
J_0(v_{\phi_\nu}, \nu)) - d_h J_1(\{v_{\phi_\tau}, \tau \}_{\partial\nu})))$, 
where the vector fields 
$v_{\phi_\nu}= v_{\phi_\nu}(\phi_\nu), v_{\phi_\tau}= v_{\phi_\tau}(\phi_\tau)$ on the corresponding fibers are defined as to induce the $J$ modified field equations (\ref{Jfieldeqs}).  

The local properties of this framework come to the forefront when the resulting structure is phrased using general boundary field theory (GBFT) language \cite{GBFT}. To each oriented codimension one surface $\Sigma$ we assign the space of first-order data over it $\Gamma_\Sigma \ni \{ \tilde{\phi} (\tau_\nu) \}_{\tau \subset \Sigma}$ (where $\nu$ is selected using $\Sigma$'s orientation), and the space is equipped with the presymplectic structure defined above, $(\Gamma_\Sigma , \omega_{L, \Sigma})$. When it happens that $\Sigma = \partial U$, the space of solutions induces a Lagrangian subspace of $(\Gamma_{\partial U} , \omega_{L, \partial U})$. The phase space $(\Gamma_{\partial U} , \omega_{L, \partial U})$ may be perfectly non degenerate. 
However, if the bulk of $U$ is used as ``propagator'' entangling degrees of freedom of the field over $\partial U$, the resulting subspace will be Lagrangian. 

Of special practical significance might be cases with $\partial U = \Sigma_f -\Sigma_i$. For them the multisymplectic formula tells us that solutions to the field equations provide an appropriate analog to a simplectomorphism (in this setting that allows for degeneracies) between $(\Gamma_\Sigma , \omega_{L, \Sigma_i})$ and  $(\Gamma_\Sigma , \omega_{L, \Sigma_f})$.

In order to make 
this short review of the framework presented in \cite{meffGBFT}
more concrete, let me 
consider the case of a nonlinear scalar field in a two-dimensional Minkowski space. 
The discretization is constructed using coordinate rectangles. 
Given an atom $\nu$, its faces will be denoted by pairs
$(\tau=0+, \tau=0-, \tau=1+, \tau=1-)$; for example, the face $\tau=0+$ is the one shared with the neighboring atom in the positive $x^0$ direction. 
The discrete action is $S_U(\phi)= \sum_U L(\tilde{\phi}(\nu))$, with 
the discrete Lagrangian being a sum of corner Lagrangians 
$L=\sum_c L^c = L^{++} + L^{+-} + L^{-+} + L^{--}$. The Lagrangian for the corner 
corresponding to increasing $x^0$ and $x^1$ directions is 
\[
L^{++}(\tilde{\phi}(\nu)) = \left\{ 
\frac{1}{2} \left[ \left(\frac{\phi_{0+} -\phi_\nu}{h}\right)^2 - 
\left(\frac{\phi_{1+} -\phi_\nu}{k}\right)^2 \right] 
+ N(\phi_\nu) \right\}hk , 
\]
where $h$ is the size of the interval between $C\nu$ and $C_{0+}$ (and $C_{0-}$) and $k$ is the size of the interval between $C\nu$ and $C_{1+}$ (and $C_{1-}$). 
The other corner Lagrangians are analogous. 

Note that the nonlinear term $N$ 
involves only 
degrees of freedom interior to the atoms; $N$ then 
enters the field equations internal to the atoms 
\[
{\rm E}_L(\cdot , \tilde{\phi}(\nu)) = 
2hk \left(
- \frac{\phi_{0+} -2\phi_\nu + \phi_{0-}}{h^2}
+ \frac{\phi_{1+} -2\phi_\nu + \phi_{1-}}{k^2}
+ 2 N'(\phi_\nu)
\right) d \phi_\nu = 0, 
\]
but the 
reader may verify that 
Cartan's form is independent of the complications of the non linear term
\[{\scriptstyle
\Theta_L(\cdot , \tilde{\phi}(1+_\nu)) = 
-2h \frac{\phi_{1+}(\nu) - \phi_\nu}{k} d\phi_{1+}
\quad , \quad 
\Theta_L(\cdot , \tilde{\phi}(0+_\nu)) = 
2k \frac{\phi_{0+}(\nu) - \phi_\nu}{h} d\phi_{0+} ,}
\]
\[{\scriptstyle
\Theta_L(\cdot , \tilde{\phi}(1-_\nu)) = 
2h \frac{\phi_\nu - \phi_{1-}(\nu)}{k} d\phi_{1-}
\quad , \quad 
\Theta_L(\cdot , \tilde{\phi}(0-_\nu)) = 
-2k \frac{\phi_\nu - \phi_{0-}(\nu)}{h} d\phi_{0-} .}
\]
This has two very appealing consequences: First, the gluing field equations $\Theta_L(\cdot , \tilde{\phi}(\tau_\nu)) = - \Theta_L(\cdot , \tilde{\phi}(\tau_{\nu'}))$ do not involve $N$, remaining simple even if the non linear term is complicated. Second, the geometric structure given by the multisymplectic form with its associated structural conservation laws (\ref{MSformula}),(\ref{thinMSformula}) is also $N$ independent, 
\[{\scriptstyle
\Omega_L(\cdot , \cdot , \tilde{\phi}(1+_\nu)) = 
-\frac{2h}{k} d\phi_\nu \wedge d\phi_{1+}
\quad , \quad 
\Omega_L(\cdot , \cdot , \tilde{\phi}(0+_\nu)) = 
\frac{2k}{h} d\phi_\nu \wedge d\phi_{0+} \quad ,}
\]
\[{\scriptstyle
\Omega_L(\cdot , \cdot , \tilde{\phi}(1-_\nu)) = 
-\frac{2h}{k} d\phi_\nu \wedge d\phi_{1-}
\quad , \quad 
\Omega_L(\cdot , \cdot , \tilde{\phi}(0-_\nu)) = 
\frac{2k}{h} d\phi_\nu \wedge d\phi_{0-} \quad .}
\]

\section{Observable currents}
\label{OCsSection}
As mentioned in the introductory section, there are strong motivations to 
study observables generated from integration on codimension one surfaces, 
$\sum_\Sigma \; \tilde{\phi}^\ast \; F$. 
Simple examples of $n-1$ cochains on $J^1Y_\Delta$ are $\iota_{\tilde{v}} \Theta_L$ and $\iota_{\tilde{w}} \iota_{\tilde{v}} \Omega_L$.

There are three features that functions of this type 
should have, {\em when evaluated on solutions}, to be called  physical observables (or complete observables). 
(i) The orientation of the integration surface $\Sigma$, specifying the side of the surface used to evaluate the integral and to keep track of signs, must be relevant only for keeping track of the signs. Equation (\ref{i}) below. 
(ii) The observable must be insensitive to local deformations of the integration surface $\Sigma$, making $\Sigma$ an auxiliary object (relevant only up to homology). 
This property ensures that more than a local function depending on first-order initial data on on $\Sigma$, it is a function depending on the solution. Equation (\ref{ii}) below. 
(iii) The observables must be independent of 
non propagating degenerate directions of the linearized field equations. 
In the next paragraphs I motivate and explain the last condition that is formally stated in equation (\ref{iii}) below.

A non propagating degenerate direction of the linearized field equations induces a 
deformation of a given solution in a spacetime localized fashion without the linearized field equations 
demanding a consequent 
deformation of the field in neighboring locations. This type of spacetime localized deformations that do not propagate cannot have an origin in the past portion of the history, nor in the character of possible boundary conditions, nor will they be relevant in other spacetime locations. Thus, they may be regarded as physically unimportant. In a discrete model this type of degenerate directions may appear as a ``lattice artifact,'' and they are not expected to be relevant in a continuum limit. An additional reason to consider  
this type of solutions of the linearized field equations as unphysical is that by doing so one is lead to a 
notion of observable current that is tightly controlled. 

After giving a more formal definition of this concept, I will show that it is equivalent to the condition for a solution of the linearized field equations to be in the null space of the multisymplectic form.

Propagation of small deformations of a solution are governed by the linearized gluing equation 
${\cal L}_{\tilde{v}} ( \Theta_L(\cdot , \tilde{\phi}(\tau_\nu)) + \Theta_L(\cdot , \tilde{\phi}(\tau_{\nu'})) )=0$. Picture a codimension one surface $\Sigma$ locally separating spacetime into two regions; 
given a solution $\phi \in {\rm Sols}_U$, 
a small deformation $\tilde{v} \in {\mathfrak F}_{\tilde{\phi}(U)}$ 
that does not propagate from one of the regions to the other is one that could be different from zero on one side and vanish on the other side. Notice that since the two regions share the degrees of freedom over $\Sigma$, the above assumption implies $\tilde{v}(\tau_\nu) = 0$ for all $\tau \subset \Sigma$. Thus, the condition for a small deformation 
to be non propagating across a codimension one surface is 
\[
0={\cal L}_{\tilde{v}} \Theta_L(\cdot , \tilde{\phi}(\tau_\nu)) = 
(\iota_{\tilde{v}}d + d \iota_{\tilde{v}}) \Theta_L(\cdot , \tilde{\phi}(\tau_\nu)) =
- \iota_{\tilde{v}} \Omega_L(\cdot , \tilde{\phi}(\tau_\nu)) . 
\]
Thus, non propagating deformations correspond to the null space of the multisymplectic form.%
\footnote{
For an off-shell history, the linearized gluing field equations involve $J^1$. The same argument implies that 
non propagating deformations are the null directions of the modified multisymplectic form 
$\Omega_L + d J_1$. 
} 

Other two important principles considered while constructing the definitions given below are the following: 
(i) The notion of physical observables will be more useful if it results from the restriction to the space of solutions of a function defined for any history. (ii) For every field theory, there should be enough physical observables as to separate points in the space of solutions.

\subsection{Observable currents and physical observables}
\label{definitions}
\begin{definition}[Observable currents]
A $n-1$ cochain $F$ on $U_F\subset J^1Y_\Delta$ is called an {\bf observable current} $F \in {\rm OC}_U$, if and only if for all 
$\phi \in {\rm Sols}_U$ such that $\tilde{\phi}(U) \subset U_F$ 
the following conditions are satisfied: 
\begin{equation}\label{i}
F(\tilde{\phi}(\tau_\nu)) \doteq F(\tau, \phi_\tau , ( \phi_\tau, \phi_\nu)) 
= - F(\tilde{\phi}(\tau_{\nu'})) \doteq F(-\tau, \phi_\tau , ( \phi_\tau , \phi_{\nu'})) . 
\end{equation}
\begin{equation}\label{ii}
\sum_{\partial U'} \tilde{\phi}^\ast \; F = 
\sum_{\tau \subset \partial U'} \; F (\tilde{\phi}(\tau_\nu))
=0 \quad \quad 
\forall \; \; U' \subset U ,	
\end{equation}
where the subindex $\nu$ in $\tau_\nu$ refers to the $n$ dimensional atom containing $\tau$ that is interior to $U'$. \\
The null directions of the multisymplectic form are also null directions of $dF$. More precisely, 
\begin{equation}\label{iii}
\tilde{\phi}^\ast \iota_{\tilde{v}} dF  = 0  
\end{equation}
for any $\tilde{v} \in {\mathfrak F}_{U_F}$ 
satisfying 
$\tilde{\phi}^\ast \iota_{\tilde{v}} \Omega_L|_{{\mathfrak F}_{U_F}} = 0$. 
\end{definition}

Clearly, the intention is to define observables by integrating observable currents. Before doing so, as I point out below, 
in general field theories 
it will be necessary to give a less stringent definition of observable currents. 
In Subsection \ref{LocalMeasurement} I give an algorithm to construct observable currents capable of measuring arbitrary local physical properties of the field; consequently, constructing observables capable of 
separating points in the space of physically distinct solutions. 
If that algorithm is applied 
with the above notion of observable currents in mind, 
one realizes that one of the necessary conditions is stronger that the condition that can be guaranteed for generic field theories. 
Below, I formulate the notion of weak observable currents for which the mentioned algorithm can be carried out. 

\begin{definition}[Weak observable currents]
Consider $[\Sigma]$, 
a given homology class of $n-1$ dimensional surfaces in the domain of interest. 
Let $F$ be a function 
of embedded oriented $n-1$ surfaces contained in 
$U_F\subset J^1Y_\Delta$ 
of the special form $\tilde{\phi}(\Sigma')$ with 
$\phi \in {\rm Sols}_U \subset {\rm Hists}_U$ and $\Sigma' \in [\Sigma]$. 
$F$ is called a 
{\bf $\Sigma$-weak observable current}, $F \in {\rm wOC}_{\Sigma, U}$, 
if for every solution $\phi \in {\rm Sols}_U$ 
such that $\tilde{\phi}(U) \subset U_F$ 
the following conditions are satisfied: 
\begin{equation}\label{wi}
F(\tilde{\phi}(\Sigma)) = - F(\tilde{\phi}(\bar{\Sigma})) . 
\end{equation}
\begin{equation}\label{wii}
F(\tilde{\phi}(\Sigma + \partial U')) = F(\tilde{\phi}(\Sigma)) 
\quad \forall \; U' \subset U. 
\end{equation}
Let $f_\Sigma : {\rm Sols}_U \to \R$ be defined as $f_\Sigma(\phi) = F(\tilde{\phi}(\Sigma))$. 
Then 
\begin{equation}\label{stratifdif}
d f_\Sigma (\phi)[v] = \sum_\Sigma \tilde{\phi}^\ast \; \iota_{\tilde{v}} dF , 
\end{equation}
with 
$dF$ a 
one-form valued $n-1$ cochain in $J^1Y_\Delta$ that is independent of $\Sigma \in [\Sigma]$. \\
%
The null directions of the multisymplectic form restricted to $\Sigma$ 
are also null directions of $dF$. More precisely, 
\begin{equation}\label{wiii}
 \sum_\Sigma \tilde{\phi}^\ast \; \iota_{\tilde{v}} dF |_{{\mathfrak F}_{U_F}}  = 0  
\end{equation}
for every 
$\tilde{v} \in {\mathfrak F}_{U_F}$ 
satisfying 
$ \sum_\Sigma \tilde{\phi}^\ast \; \iota_{\tilde{v}} \Omega_L|_{{\mathfrak F}_{U_F}} = 
\iota_{\tilde{v}} \omega_{L, \Sigma}|_{{\mathfrak F}_{U_F}} = 0$. 
\end{definition}
Notice that the spaces  ${\rm OC}_U$, ${\rm wOC}_{\Sigma, U}$ 
are vector spaces, and that observable currents induce weak observable currents.

\begin{definition}[Observables]
An {\bf observable} is the result of 
``integrating'' an observable current on a codimension one oriented surface. 
More precisely it is an evaluation of a weak observable current that, in the case of 
an observable current, may be calculated as a sum 
\[
f_\Sigma(\phi) \doteq F(\tilde{\phi}(\Sigma)) \overset{\tiny {\rm OC}_U}{=} 
\sum_\Sigma \; 
\tilde{\phi}^\ast \; F . 
\]
\end{definition}

A simple consequence of 
the defining properties of (weak) observable currents is that for $\phi \in {\rm Sols}_U$
\begin{enumerate}
\item
$f_{\bar{\Sigma}}(\phi) = - f_\Sigma(\phi)$ 
\item
$\Sigma' = \Sigma + \partial U'$ for some $U' \subset U$ implies $f_{\Sigma'}(\phi) = f_\Sigma(\phi)$
\item
$v$ in the null space of $\bar{\omega}_{L, \Sigma}$ implies ${\cal L}_v f_\Sigma(\phi)= 0$
\end{enumerate}

{\em Remarks}: 
\begin{itemize}
\item
An observable current $F \in {\rm OC}_U$ is a {\bf local} object, 
and it induces observables that are functions on ${\rm Hists}_U$.  
Examples are presented in some detail in Subsection \ref{examples}. 
One family is composed by Noether currents 
$N_\xi = -\iota_{\tilde{v}_\xi} \Theta_L$, where $\Theta_L$ was defined before in terms of partial derivatives of the Lagrangian and $\tilde{v}_\xi$ is the vector field  induced by symmetry generator $\xi\in Lie({\cal G})$. Both, 
$\Theta_L$ and $\tilde{v}_\xi$ can be evaluated anywhere in the jet bundle; 
once a symmetry generator $\xi\in Lie({\cal G})$ 
has been chosen, one can proceed to evaluate $N_\xi$ anywhere in the bundle. 
A much larger family of examples is composed by symplectic products $F=\iota_{\tilde{w}} \iota_{\tilde{v}} \Omega_L$, 
where $\Omega_L = - d \Theta_L$ involves partial derivatives of the Lagrangian and 
${\tilde{v}}, {\tilde{w}} \in {\mathfrak F}_{\tilde{\phi}(U)} \subset {\mathfrak X}(J^1 Y_\Delta |_U)$ are two solutions of the linearized field equations (\ref{linearizedFE}). The ingredients of the observable current: $\Omega_L$ and ${\tilde{v}}, {\tilde{w}}$ can be evaluated anywhere inside a neighborhood of $\tilde{\phi}(U) \subset J^1 Y_\Delta |_U$; 
once two solutions of the linearized field equations ${\tilde{v}}, {\tilde{w}} \in {\mathfrak F}_{\tilde{\phi}(U)}$ 
have been chosen, one can proceed to evaluate $F$ anywhere in the mentioned neighborhood. 
\item
A weak observable current $F \in {\rm wOC}_U$ is in general {\bf non-local}, 
and it induces observables that are functions on ${\rm Sols}_U \subset {\rm Hists}_U$. 
First, it acts on whole $n-1$ dimensional surfaces 
of the type $\tilde{\phi}(\Sigma')$ 
and in general cannot be calculated as an integral of a cochain. 
Second, by an abuse of notation I wrote the differential of the induced observable as an integral using the symbol $dF$, but this one-form valued $n-1$ cochain in $J^1Y_\Delta$ is not 
required to be closed. 
%
%
%
In Subsection \ref{LocalMeasurement} I 
give an algorithm to construct weak observable currents in which the properties of 
$dF$ are dictated by the linearized field equations. 
Propagation according to those equations guarantees that 
if $\sum_\Sigma \tilde{\phi}^\ast \; dF|_{{\mathfrak F}_{U_F}}$ is closed, so is the form constructed by a similar integral on $\Sigma + \partial U'$. 
\item
The properties of observable currents make it natural to define their ``integration'' on spacetime $n-1$ cycles modulo boundaries. 
\end{itemize}

Before closing this subsection, containing the main definitions of this work, I point out that {\em in the case of gauge theories and extended gauge theories} 
requirement (\ref{iii}) in the definition of observable currents should be strengthened. 
Apart from demanding that observable currents be independent of null directions of the multisymplectic form, also independence of any direction in which 
$\tilde{v}$ corresponds to a gauge symmetry should be part of the definition. In that way, observable currents model physical gauge invariant observables. 
Thus, the notion of observable current applies to scalar fields, non linear sigma models, gauge fields and extended gauge fields (including general relativity). For a thorough explanation of the multisymplectic framework over a discretized spacetime for these types of field theories, see \cite{meffGBFT}.

%

\subsection{Two families of observable currents}
\label{examples}

\subsubsection*{Observable currents from the symplectic product} 

The $n-1$ cochain \hskip0.3cm 
$F=\iota_{\tilde{w}} \iota_{\tilde{v}} \Omega_L$ \\

\noindent
is an observable current 
$F \in {\rm OC}_U$ 
(in a vicinity of $\tilde{\phi}(U)$ for some $\phi \in {\rm Sols}_U$) 
if the vector fields are first variations, ${\tilde{v}}, {\tilde{w}}\in {\mathfrak F}_{\tilde{\phi}(U)}$. 


The physical interpretation of this observable current will be easier to explain after the following sections. Here is a preliminary explanation in a context in which there is a codimension one surface $\Sigma$ that in the interior of the domain of interest has no boundary, $\partial \Sigma \subset \partial U$. By integration on this surface, the multisymplectic form 
induces a presymplectic structure in the 
space of solutions, $(T {\rm Sols}_U, \bar{\omega}_{L, \Sigma})$. 
Consider a situation in which the first variations mentioned above satisfy the conditions 
$\sum_\Sigma \tilde{\phi}^\ast {\cal L}_{\tilde{v}} \Omega_L=0$, 
$\sum_\Sigma \tilde{\phi}^\ast {\cal L}_{\tilde{w}} \Omega_L=0$. Then, the first variations  induce locally Hamiltonian vector fields 
$v, w$ in the presymplectic space $(T {\rm Sols}_U, \bar{\omega}_{L, \Sigma})$. These vector fields are 
such that 
$f_\Sigma(\phi)= \sum_\Sigma \; 
\tilde{\phi}^\ast \; F = \iota_w \iota_v \bar{\omega}_{L, \Sigma}$. In other words, 
$f_\Sigma$ is the symplectic product of the mentioned locally Hamiltonian vector fields; additionally, 
the Hamiltonian vector field of the resulting observable is the commutator of the vector fields 
$d f_\Sigma = -\iota_{[ v, w]} \bar{\omega}_{L, \Sigma}$. 
It is important to notice that the multisymplectic formula guarantees that if 
$\sum_\Sigma \tilde{\phi}^\ast {\cal L}_{\tilde{v}} \Omega_L|_{{\mathfrak F}_{\tilde{\phi}(U)}}=0$, then the same property holds when the integral is over any homologous surface $\Sigma' = \Sigma + \partial U'$. 

The spacetime covariant version of these clues is that 
\begin{equation}\label{essentiallyHamiltonian}
d F = -\iota_{[ \tilde{v}, \tilde{w} ]} \Omega_L 
+ \iota_{\tilde{v}} {\cal L}_{\tilde{w}} \Omega_L 
- \iota_{\tilde{w}}  {\cal L}_{\tilde{v}} \Omega_L . 
\end{equation}
Therefore, $F$ does not exactly satisfy the equation linking it to a Hamiltonian vector field, but 
its local failure to do so vanishes when integrated on $n-1$ surfaces $\tilde{\phi}(\Sigma')$ 
whose spacetime projection $\Sigma'$ is in the homology class $[\Sigma]$ and 
has no boundary in the interior of $U$, $\partial \Sigma'|_U \subset \partial U$. 
In the next subsection, I will introduce a notion of Hamiltonian observable currents which is sufficiently ample to  
contain the present family of observable currents. 

This will be a key family of examples to have in mind during the rest of the paper. 
A central question will be: {\em ``are all local properties of the field measurable with this family of observable currents?''} 

Before passing to the next example, I would like to remark that observables of this type are 
defined only implicitly because their ingredients are vector fields satisfying 
local conditions -- in particular  
the linearized field equations (\ref{linearizedFE}). One would have to exhibit explicit solutions of those equations to explicitly construct the desired observable currents. On the other hand, observable currents of this type exist in all situations and provide local information about the field.

\subsubsection*{Observable currents from symmetries}

In theories with symmetries, Noether's theorem yields another family of observable currents. 
In contrast to the previous family of observable currents, there are non trivial observable currents of this type only when 
the action of the field theory has symmetries. But, on the other hand, formulas for this type of observable currents will be written explicitly.

Let a Lie group ${\cal G}$ act on the standard fiber 
${\cal F}$; the diagonal action of ${\cal G}$ 
on $Y_\Delta$, 
written in first order format, is 
\[
(g\tilde{\phi})(\nu) = (\nu, g(\phi_\nu) , \{ g(\phi_\tau)\}_{\tau \subset \partial \nu}). 
\]

Now consider the case in which this action  
leaves the Lagrangian invariant: 
$L(g\tilde{\phi}(\nu)) = L(\tilde{\phi}(\nu))$ for all atoms $\nu$ in $U$, for any first-order history 
$\tilde{\phi}$ and for any $g\in {\cal G}$. 
In this case, the ${\cal G}$-action preserves 
$S$ and the subspace of extrema. 
For any  
$\xi\in Lie({\cal G})$ the corresponding generator 
of the ${\cal G}$-action on $J^1Y_\Delta$ 
is a vector field $\tilde{v}_\xi$, 
which induces a variation 
that is in the kernel of $dS$ evaluated on any history; additionally, if the considered history is a 
solution $\tilde{\phi} \in {\rm Sols}_U$ the induced vector field is a first variation 
$\tilde{v}_\xi \in {\mathfrak F}_{\tilde{\phi}(U)}$. 
Let us define an 
$n-1$ cochain on $J^1Y_\Delta$ associated with each 
$\xi\in Lie({\cal G})$ --the Noether current-- 
\[
N_\xi = -\iota_{\tilde{v}_\xi} \Theta_L . 
\]
Noether's theorem states that Noether's current is 
indeed an observable current 
$N_\xi \in {\rm OC}_U$. The proof that it satisfies 
condition (\ref{i}) follows directly from 
the gluing field equations (\ref{gluingFEQ}); the proof that it satisfies 
condition (\ref{ii}) follows from 
\[
0 = dS(\tilde{\phi}) [v_\xi] = 
\sum_{\partial U} \tilde{\phi}^\ast(
\iota_{\tilde{v}_\xi} \Theta_L) , 
\]
and the proof that it satisfies 
condition (\ref{iii}) follows from 
\[
dN_\xi = -\iota_{\tilde{v}_\xi} \Omega_L , 
\]
which in turn follows from Cartan's identity 
$L_{\tilde{v}_\xi}= \iota_{\tilde{v}_\xi}d + d\iota_{\tilde{v}_\xi}$. 
The last equation relates the current $N_\xi$ to a Hamiltonian vector field 
$\tilde{v}_\xi$. Observable currents that participate in an equation of this type 
will be called strict Hamiltonian observable currents. 
The primary focus of our attention in the following sections will be Hamiltonian vector fields and the related observable currents.

Noether's theorem also talks about the algebraic structure 
that the corresponding observables acquire with Poisson's bracket. In this paper 
I will present this extra structures in Subsection \ref{ssPoisson}. 

A construction leading to examples of weak observable currents will be given after the next subsection. 

\subsection{Locally Hamiltonian vector fields and\\
observable currents}
\label{LocHam}
%
%
%
%

Noether currents satisfy an equation of the type $dF=-\iota_{\tilde{v}} \Omega_L$ . 
In the first family of examples of observable currents, a weaker equation appeared instead (\ref{essentiallyHamiltonian}). 
Equations of this type are natural candidates to be 
the origin, within the multisymplectic framework, of 
the famous equation relating Hamiltonian vector fields with their corresponding observables. 
Below I will show a realization of this expectation. 

%
%
%

\begin{definition}[Strict Hamiltonian observable currents and \\
strict Hamiltonian vector fields]
An observable current $F$ and a first variation $\tilde{v} \in {\mathfrak F}_{U_F}$
that (inside a neighborhood $U_F\subset J^1Y_\Delta$) 
participate in the equation 
\[
dF=-\iota_{\tilde{v}} \Omega_L
\]
will be called, respectively, 
{\bf strict Hamiltonian observable current} $F \in {\rm sHOC}_U \subset {\rm OC}_U$ and 
{\bf strict Hamiltonian vector field} 
$\tilde{v} \in {\mathfrak F}_{U_F}^{\rm sH} \subset {\mathfrak F}_{U_F}$.  
\end{definition}

Two Hamiltonian vector fields related to the same observable current differ by a vector field that is in the 
null space of $\Omega_L$. 
Thus, the above equation relates Hamiltonian observable currents 
and classes of Hamiltonian vector fields. 


It is interesting to study the local condition that makes it possible for first variations to be 
strict Hamiltonian vector fields. This happens when the differential form $\sigma_{\tilde{v}} \doteq -\iota_{\tilde{v}} \Omega_L$ is closed. 
Since the multisymplectic form is closed, this condition is equivalent to ${\cal L}_{\tilde{v}} \Omega_L = 0$. 
Thus, it is necessary to study the compatibility of ${\cal L}_{\tilde{v}} \Omega_L = 0$ with equation (\ref{linearizedFE}), characterizing first variations. 
It turns out that the two requirements just mentioned are compatible only in very special situations. 
However, the multisymplectic formula implies that the field equations are compatible with 
imposing that for $\phi \in {\rm Sols}_U$ the differential form written below is closed 
\[
\sum_\Sigma \tilde{\phi}^\ast \; \sigma_{\tilde{v}}|_{{\mathfrak F}_{U_F}} = 
- \sum_\Sigma \tilde{\phi}^\ast \;  \iota_{\tilde{v}} \Omega_L |_{{\mathfrak F}_{U_F}} . 
\]
More precisely, if the vector field is a first variation and the differential form written above is closed, then any 
deformation of the integration surface within its homology class maintains the resulting form closed.%
\footnote{
For $\phi \in {\rm Hists}_U^J$, the relevant form is 
$\sum_\Sigma \tilde{\phi}^\ast \; \sigma_{\tilde{v}}^J|_{{\mathfrak F}_{U_F}^J} =
- \sum_\Sigma \tilde{\phi}^\ast \;  \iota_{\tilde{v}} 
(\Omega_L + dJ_1)|_{{\mathfrak F}_{U_F}^J}$. 
} 
A weaker version of locally Hamiltonian vector fields could be defined requiring that this form be closed. 
However, it is not appealing to define a class of intrinsically local objects, like vector fields, by means of the properties of an integrated expression. 
Below I give a formulation of Hamiltonian observable currents that is based on this idea, and leads to a local condition that is weaker than the expression explored above. 
\begin{definition}[Hamiltonian observable currents and Hamiltonian vector fields]
%
Consider $[\Sigma]$, 
a given homology class of $n-1$ dimensional surfaces without boundary in the domain of interest. 
An observable current $F \in {\rm OC}_{\Sigma, U}$ and a first variation $\tilde{v} \in {\mathfrak F}_{U_F}$ 
(defined in a neighborhood $U_F\subset J^1Y_\Delta$) 
will be called, respectively, 
{\bf $\Sigma$-Hamiltonian observable current} $F \in {\rm HOC}_{\Sigma, U}\subset {\rm OC}_{\Sigma, U}$ and 
{\bf $\Sigma$-Hamiltonian vector field} $\tilde{v} \in {\mathfrak F}_{U_F}^{\Sigma{\rm H}} \subset {\mathfrak F}_{U_F}$ 
if they participate in the equation 
\[
dF=-\iota_{\tilde{v}} \Omega_L + \alpha_{\tiny \Sigma} ,
\]
where $\alpha_{\tiny \Sigma}$ is such that for any $\phi \in {\rm Sols}_U$ 
\[
d_h \alpha_{\tiny \Sigma} |_{{\mathfrak F}_{U_F}} = 0 \quad \mbox{ and } \quad
\sum_{\Sigma} \tilde{\phi}^\ast \; \alpha_{\tiny \Sigma} |_{{\mathfrak F}_{U_F}} = 0 . 
\]
\end{definition}

The following condition characterizes first variations that are $\Sigma$-Hamiltonian vector fields 
in a neighborhood of $U_{\tilde{v}}\subset J^1Y_\Delta |_U$. 
\begin{definition}[Locally Hamiltonian vector fields]
A first variation $\tilde{v}$ 
(defined in a neighborhood $U_{\tilde{v}}\subset J^1Y_\Delta |_U$) 
is said to be {\bf $\Sigma$-locally Hamiltonian} 
$\tilde{v} \in {\mathfrak F}_{U_{\tilde{v}}}^{\Sigma{\rm LH}} \subset {\mathfrak X}(U_{\tilde{v}}\subset J^1Y_\Delta |_U)$, 
if and only if for every $\phi \in {\rm Sols}_U$ 
\begin{equation}
\label{sigmaLHVF}
\sum_\Sigma \tilde{\phi}^\ast \; 
{\cal L}_{\tilde{v}} \Omega_L |_{{\mathfrak F}_{U_{\tilde{v}}}} = 
{\cal L}_{\tilde{v}} \omega_{L, \Sigma} |_{{\mathfrak F}_{U_{\tilde{v}}}} = 
0  \quad \mbox{ for } \Sigma \in [\Sigma ] .  
\end{equation}
A first variation $\tilde{v}$ that is $\Sigma$-locally Hamiltonian 
for any $\Sigma$ is simply called {\bf locally Hamiltonian} 
$\tilde{v}\in{\mathfrak F}_{U_{\tilde{v}}}^{\rm LH} \subset {\mathfrak X}(U_{\tilde{v}}\subset J^1Y_\Delta |_U)$. 
\end{definition}
A simple consequence of Cartan's identity and $\omega_{L, \Sigma}$ 
(or $\Omega_L$) being closed is that 
first variations $\tilde{v} \in {\mathfrak F}_{\tilde{\phi}(U)}$ 
that are in the null space of $\omega_{L, \Sigma}$ 
(or $\Omega_L$) are $\Sigma$-locally Hamiltonian 
$\tilde{v} \in {\mathfrak F}_{U_{\tilde{v}}}^{\Sigma{\rm LH}}$. 
Then, it is mathematically possible to consider equivalence classes of locally Hamiltonian vector fields up to the null space of the appropriate form. 
The physical meaning of these equivalence classes depends on the details of the circumstance.

Consider observables of the type $F=\iota_{\tilde{w}} \iota_{\tilde{v}} \Omega_L$. Equation (\ref{essentiallyHamiltonian}) 
showed that its differential is 
$d F = -\iota_{[ \tilde{v}, \tilde{w} ]} \Omega_L 
+ \iota_{\tilde{v}} {\cal L}_{\tilde{w}} \Omega_L - \iota_{\tilde{w}}  {\cal L}_{\tilde{v}} \Omega_L $. 
Thus, the family of observable currents formed by symplectic products of first variations that are 
$\Sigma$-locally Hamiltonian, 
$\tilde{v}, \tilde{w} \in {\mathfrak F}_{U_{\tilde{v}}}^{\Sigma{\rm LH}}$, 
is a family of $\Sigma$-Hamiltonian observable currents. The corresponding 
$\Sigma$-Hamiltonian vector field is $[ \tilde{v}, \tilde{w} ]$, 
and the residual field is $\alpha_{\tiny \Sigma}= \iota_{\tilde{v}} {\cal L}_{\tilde{w}} \Omega_L 
- \iota_{\tilde{w}}  {\cal L}_{\tilde{v}} \Omega_L $. 

A Noether current $N_\xi$ is 
a strict Hamiltonian observable current, and its corresponding strict Hamiltonian vector field is $\tilde{v}_\xi$.

In classical mechanics, when the structural form is symplectic instead of merely presymplectic, every function on phase space has an associated hamiltonian vector field. 
Here $\Omega_L$ may be degenerate, but by definition only currents whose differential has the same null directions as the multisymplectic form are considered; thus, a direct consequence of the definition is that $\Omega_L$, as a local map from (equivalence classes of) 
vectors to one-forms with the given null space, {\em is invertible}. 
Thus, one may wonder if all observable currents are Hamiltonian in an appropriate sense. 
Here is an example that in field theory this is not the case. An observable current 
$F=\iota_{\tilde{w}} \iota_{\tilde{v}} \Omega_L$ for which 
$\sum_\Sigma \tilde{\phi}^\ast {\cal L}_{\tilde{v}} \Omega_L=0$ and 
$\sum_\Sigma \tilde{\phi}^\ast {\cal L}_{\tilde{w}} \Omega_L=0$ are not both satisfied for the same $[\Sigma]$ is not 
Hamiltonian according to any of the definitions given above.


Now I define a corresponding notion for weak observable currents. 

\begin{definition}[Weak Hamiltonian observable currents and Hamiltonian vector fields]
%
Consider $[\Sigma]$, 
a given homology class of $n-1$ dimensional surfaces without boundary in the domain of interest. 
A weak observable current $F \in {\rm wOC}_{\Sigma, U}$ and a 
vector field $\tilde{v} \in {\mathfrak F}_{U_F}$ 
(defined in a neighborhood $U_F\subset J^1Y_\Delta$) 
will be called, respectively, 
{\bf $\Sigma$-Weak Hamiltonian observable current} $F \in {\rm wHOC}_{\Sigma, U} \subset {\rm wOC}_{\Sigma, U}$ and 
{\bf $\Sigma$-Hamiltonian vector field} 
$\tilde{v} \in {\mathfrak F}_{U_F}^{\Sigma{\rm wH}} \subset {\mathfrak F}_{U_F}$ 
if for every solution $\phi \in {\rm Sols}_U$ 
they participate in the equation 
\[
\tilde{\phi}^\ast \; dF|_{{\mathfrak F}_{U_F}}=-\tilde{\phi}^\ast \; (\iota_{\tilde{v}} \Omega_L |_{{\mathfrak F}_{U_F}} 
+ \alpha_{\tiny \Sigma}|_{{\mathfrak F}_{U_F}}) ,
\]
where $\alpha_{\tiny \Sigma}$ is such that 
\[
\tilde{\phi}^\ast \; d_h \alpha_{\tiny \Sigma} |_{{\mathfrak F}_{U_F}} = 0 \quad \mbox{ and } \quad
\sum_{\Sigma} \tilde{\phi}^\ast \; \alpha_{\tiny \Sigma} |_{{\mathfrak F}_{U_F}} = 0 . 
\]
\end{definition}

The notion of $\Sigma$-locally Hamiltonian vector fields given above, 
also applies in the context of weak Hamiltonian observable currents. 
In the next subsection,
the notion of locally Hamiltonian vector field is used to construct weak observable currents.

\subsection{Construction of weak observable currents}\label{ConstrWocs}

In this subsection I give an implicit construction of weak observable currents. 
Consider $[\Sigma]$, a given homology class of $n-1$ dimensional surfaces embedded in the domain of interest. 
Let $\tilde{v} \in {\mathfrak F}_{U_{\tilde{v}}}^{\Sigma{\rm LH}}$ be a locally Hamiltonian first variation and define 
$\sigma_{\tilde{v}} \doteq -\iota_{\tilde{v}} \Omega_L$. 
By construction, it is clear that for $\phi \in {\rm Sols}_U$ and $\Sigma \in [\Sigma]$ 
\[
d\sum_\Sigma \tilde{\phi}^\ast \; 
\sigma_{\tilde{v}} |_{{\mathfrak F}_{U_{\tilde{v}}}}= 
\sum_\Sigma \tilde{\phi}^\ast \; 
d \sigma_{\tilde{v}}|_{{\mathfrak F}_{U_{\tilde{v}}}} = 0 . 
\]
Thus, $\sum_\Sigma \tilde{\phi}^\ast \; \sigma_{\tilde{v}}$ is the differential of a local function on the space of solutions. 
This means that in a neighborhood of ${\rm Sols}_U$ there is a function $f_\Sigma$ determined by its differential 
$d f_\Sigma (\phi)[v] = \sum_\Sigma \tilde{\phi}^\ast \; \iota_{\tilde{v}} \sigma_{\tilde{v}}$ 
and an integration constant $k_\Sigma= f_\Sigma (\phi_0)$. 
The objective is to construct a weak observable current from this function by setting 
\[
F(\tilde{\phi}(\Sigma))= f_\Sigma(\tilde{\phi}) . 
\]
Then, the conditions $F(\tilde{\phi}(\bar{\Sigma})) = - F(\tilde{\phi}(\Sigma))$, 
$F(\tilde{\phi}(\Sigma + \partial U')) = F(\tilde{\phi}(\Sigma))$ required for weak observable currents induce requirements on the differential (which are satisfied by construction) and on the $\Sigma$-dependent integration constant $k_\Sigma$. 
The conditions are  
\[
k_{\bar{\Sigma}} = - k_\Sigma , \quad k_{\Sigma + \partial U'} = k_\Sigma . 
\]
Thus, a $\Sigma$-weak observable current $F_{\tilde{v} k}\in {\rm wHOC}_{\Sigma, U}$ 
is determined by a $\Sigma$ locally Hamiltonian vector field 
$\tilde{v} \in {\mathfrak F}_{U_{\tilde{v}}}^{\Sigma{\rm LH}}$ and an integration constant $k$ that is independent of the element on the homology class $[\Sigma ]$, and which changes sign when the orientation of the surface is reversed. 
The construction guarantees that 
$F_{\tilde{v} k}$ is a Hamiltonian weak observable current with Hamiltonian vector field $\tilde{v}$.

It is worth mentioning that the constructed weak Hamiltonian observable current, is special because its residual form $\alpha_\Sigma$ vanishes 
$dF_{\tilde{v}, k}=-\iota_{\tilde{v}} \Omega_L$. 

A related remark that shows the relevance of observable currents is the following: 
{\em $F_{\tilde{v}, k}$ is not, in general, induced by an observable current; however, in the case that 
$\tilde{v} = [\tilde{x} , \tilde{y} ]$, 
for locally Hamiltonian vector fields $\tilde{x}, \tilde{y} \in {\mathfrak F}_{U_{\tilde{x}} \cap U_{\tilde{y}}}^{\Sigma{\rm LH}}$, 
the observable current $\iota_{\tilde{y}} \iota_{\tilde{x}} \Omega_L$ 
would have the same associated Hamiltonian vector field and consequently, the same differential in the space of solutions. 
Thus, the large class of the weak observable currents for which 
their defining locally Hamiltonian vector field 
$\tilde{v} \in {\mathfrak F}_{U_{\tilde{v}}}^{\Sigma{\rm LH}}$ is a commutator 
(or a linear combination of commutators) 
can be ``improved'' to become local.} 
The weak observable current $F_{[\tilde{x} , \tilde{y} ], k}\in {\rm wHOC}_{\Sigma, U}$ 
induces the same physical observable after integrating on $\Sigma$ 
as the observable current 
\[
\iota_{\tilde{y}} \iota_{\tilde{x}} \Omega_L + K \in {\rm OC}_{\Sigma, U} , 
\]
where $K$ is a system of local integration constants $K_\tau= K(\tilde{\phi_0}(\tau_\nu))  \in \R$ such that 
$\sum_\Sigma \tilde{\phi_0}^\ast \; (\iota_{\tilde{y}} \iota_{\tilde{x}} \Omega_L + K) = k_\Sigma$. 
Since $k_\Sigma$ changes sign when the orientation of the cell is reversed and 
depends only on the homology class of the oriented surface, the system of local integration constants is 
a closed $n-1$ cochain. 
This system $K$ of local integration constants is not unique, but 
since its ``integration'' on surfaces without boundary is fixed by the requirement of matching the given observable, 
any two systems that generate the same 
observable differ by a $n-1$ cochain that vanishes on cycles. 

If for a given field theory all 
first variations 
are commutators, then every weak observable current 
can be promoted to an observable current.

\subsection{OCs separate distinct neighboring solutions}

In Subsection \ref{definitions} I defined physical observables from observable currents. How interesting is this class of physical observables? 
Noether's theorem yields observable currents corresponding to conserved quantities associated with symmetries of the system. I also briefly showed a family of observable currents corresponding to the symplectic product of first variations. 
Finally, in the previous subsection I showed how to construct examples of weak observable currents. 
In this subsection I show that in a precise sense weak observable currents are capable of separating neighboring (physically distinct) solutions. 

Given a solution $\phi \in {\rm Sols}_U$, 
the local question under study is 
whether 
the vector space ${\rm wOC}_{\Sigma, U}$ is large enough 
to resolve the space of first variations 
$T_\phi {\rm Sols}_U$
up to directions corresponding to 
vector fields in the null space of $\Omega_L$. 

Consider a curve of solutions $\gamma(s) \in {\rm Sols}_U$ with 
$\gamma(0) = \phi \in {\rm Sols}_U$, 
$\dot{\gamma}(0) = w \in T_\phi {\rm Sols}_U$ consistent with the vector field 
$\tilde{w}\in {\mathfrak F}_{\tilde{\phi}(U)}$
that is not in the null space of $\Omega_L$. 
Since $\tilde{w}$ is a propagating first variation, it may be reasonable to 
further assume that $\tilde{w}$ is not in the null space of $\omega_{L, \Sigma}$ for at least one $n-1$ dimensional surface $\Sigma$ with $\partial \Sigma \subset \partial U$.

A given weak observable current $F \in {\rm wOC}_{\Sigma, U}$ defines the observable $f_\Sigma$. A simple question is if 
$f_\Sigma$ is capable of separating $\tilde{\phi}$ from nearby solutions in $\gamma$. 
The separability condition is
\begin{equation}\label{dOC}
\frac{d}{ds}|_{s=0} f_\Sigma(\gamma(s)) = 
\sum_\Sigma \tilde{\phi}^\ast \; dF [\tilde{w}] \neq 0 .
\end{equation}

When restricting attention to 
the family of $\Sigma$-Hamiltonian observable currents 
constructed in the previous subsection $\{ F_{\tilde{v}, k} \}$, 
the corresponding local separability condition is whether 
for $\tilde{w}\in {\mathfrak F}_{\tilde{\phi}(U)}$
which is not in the null space of $\Omega_L$ 
there is a locally Hamiltonian vector field 
$\tilde{v} \in {\mathfrak F}_{U_{\tilde{v}}}^{\Sigma{\rm LH}}$ such that 
\[
\sum_\Sigma \tilde{\phi}^\ast \; 
\iota_{\tilde{w}} \iota_{\tilde{v}} \Omega_L = 
\iota_{\tilde{w}} \iota_{\tilde{v}} \omega_{L, \Sigma}
\neq 0 . 
\]
Thus, if the assumptions stated above hold, 
weak observable currents are capable of separating points in the space of physically distinct solutions. Moreover, the proof given above shows that 
Hamiltonian observable currents are sufficient to distinguish physically distinct solutions. 

Recall that 
one of the conditions in the definition of $\Sigma$-weak observable currents 
is that they be independent of the null directions of $\omega_{L, \Sigma}$. 
Since in the case of $\Sigma$-Hamiltonian observable currents, $\omega_{L, \Sigma}$ itself is the operator converting locally Hamiltonian vector fields into differentials, it is also true that any differential whose kernel contains that 
of $\omega_{L, \Sigma}$ can be constructed using $\omega_{L, \Sigma}$. Thus, any 
$\Sigma$-weak observable current is Hamiltonian.

Before closing this section, it is worth reminding that, 
as mentioned at the end of Subsection \ref{ConstrWocs}, the constructed 
$F_{\tilde{v}, K}$ can be  ``improved'' to become local -- to become an observable current -- if $\tilde{v}$ is a commutator (or a linear combination of commutators) of 
locally Hamiltonian vector fields. 
I mentioned that if in a given field theory all first variations, 
then any weak observable current can be promoted to become local. 
It is also possible that not all weak observable currents are replaceable by observable currents, but that the separability condition 
(\ref{dOC}) is satisfied using only observable currents. This happens for field theories in which  
every first variation that is presymplecticaly orthogonal to the space generated by commutators 
of locally Hamiltonian first variations 
is necessarily in the null space of the presymplectic form. 
A study towards characterizing the field theories that have this property is underway. 

%

\subsection{Poisson brackets}\label{ssPoisson}
%

I recall that in the definition of observable currents there is no mention of any surface embedded in spacetime, and that in the definition of 
$\Sigma$-weak observable currents, 
$\Sigma$-Hamiltonian observable currents, etc, 
only the homology class of $\Sigma$ is relevant. Observable currents are spacetime currents and are not tied to a given embedded surface. 

Consider two observable currents, or weak observable currents, $F, G$ 
that are $\Sigma$-Hamiltonian with corresponding Hamiltonian vector fields 
$\tilde{v} \in {\mathfrak F}_{U_F}^{\Sigma{\rm H}}$, $\tilde{w} \in {\mathfrak F}_{U_G}^{\Sigma{\rm H}}$. 
Their Poisson bracket is the observable current $\{ F, G \}$ defined on $U_F \cap U_G$ by 
\[
\{ F, G \} (\tilde{\phi}(\tau_\nu)) 
= 
\Omega_L (\tilde{w} , \tilde{v} , \tilde{\phi}(\tau_\nu)) , 
\]
which as shown in Subsection \ref{examples}, is a $\Sigma$-Hamiltonian observable current with associated Hamiltonian vector field 
$[ \tilde{v} , \tilde{w} ]$ and residual form $\alpha_\Sigma =  \iota_{\tilde{v}} {\cal L}_{\tilde{w}} \Omega_L 
- \iota_{\tilde{w}}  {\cal L}_{\tilde{v}} \Omega_L $. 
Even if $F$ and $G$ are only weak observable currents, which are not spacetime local and are defined only on solutions, their bracket is an observable current  -- a spacetime local object defined for every history. 

It is important to notice that the definition of the bracket among observable currents is spacetime local. 

When the Poisson bracket is integrated on a codimension one embedded surface $\Sigma$, the result is the physical observable 
\[
\{ F, G \}_\Sigma (\phi)= \omega_{L, \Sigma} (\tilde{w} , \tilde{v}), 
\]
where in the right hand side of this equation the dependence on the field is hidden in the definition of $\Sigma$'s presymplectic form 
$\omega_{L, \Sigma} (\tilde{v} , \tilde{w} ) = 
\sum_\Sigma \tilde{\phi}^\ast \; \iota_{\tilde{w}} \iota_{\tilde{v}} \Omega_L$. 

With the bracket defined above, the space of observable currents ${\rm OC}_{\Sigma, U}$ 
and the space of weak observable currents ${\rm wOC}_{\Sigma, U}$
become Lie algebras. Bilinearity and anticommutativity 
of the bracket follow directly from the corresponding properties of $\Omega_L$. The Jacobi relation deserves more attention. 
I mentioned above that the Hamiltonian vector field of the bracket is the commutator of the corresponding vector fields. This property together with the Jacobi identity for the commutator of vector fields 
implies that the Hamiltonian vector field of $J_{FGH}= \{\{ F, G \} , H \} +  \{\{ H, F \} , G \} +  \{\{ G, H \} , F \}$ is zero. Additionally, in contrast with related proposals in the continuum \cite{MSobs, DynObsEtc} 
(where the Jacobi relation is only valid up to an exact form), for the bracket defined above the Jacobi relation holds 
\[
\{\{ F, G \} , H \} +  \{\{ H, F \} , G \} +  \{\{ G, H \} , F \} = 0 . 
\]
The reason behind this fact will show some differences of this discrete framework for field theory and the continuum. 
The evaluation of the observable current $J_{FGH}$ on an elementary $n-1$ chain $\tilde{\phi}(\tau_\nu)$ is a calculation based on the two-form $\Omega_L (\cdot , \cdot , \tilde{\phi}(\tau_\nu))$ that is a closed two-form. Thus, 
for any surface $\Sigma$ the evaluation of 
$\sum_\Sigma \tilde{\phi}^\ast J_{FGH}$ is a sum of terms corresponding to the cells in the triangulation of $\Sigma$ in which each term vanishes. One discovers that this discrete formalism for field theory is set up in a discrete version of the jet bundle that can be decomposed into a collection of phase spaces with presymplectic forms (which are tightly coordinated to work together). 

In addition to being a Lie algebra, 
the product  among observable currents defined below converts this structure into a Poisson algebra: 
\[
F \cdot G \; (\tilde{\phi}(\tau_\nu))=  F(\tilde{\phi}(\tau_\nu)) G(\tilde{\phi}(\tau_\nu)). 
\]
With this product the bracket satisfies the Leibniz rule because, when restricted to any elementary $n-1$ chain $\tilde{\phi}(\tau_\nu)$, the product is the ordinary product of functions on the corresponding phase space 
with an algebra of functions which is a Poisson algebra. Kanatchikov defined a product among $k$-forms on his polysymplectic formalism \cite{Kanatchikov}. His product is designed in such a way that the product among two $n$-forms yields another $n$-form. The product defined above can be seen as a modification of Kanatchikov's product that stabilizes $n-1$ forms. A continuum analog of this product would use a volume form on codimension one surfaces to define an appropriate Hodge star operator.

\subsection{Localized measurement}
\label{LocalMeasurement}

Observable currents and weak observable currents can be used to measure physically relevant local properties of the field. 
When one thinks of a spacetime localized measurement, the premise is that there is an interaction of the measuring device and the system under study that is spacetime localized, and any effect of the measurement in other spacetime locations is caused by propagation of the direct disturbance as commanded by the field equations. {\em (Weak) observable currents do not know of this distribution of responsibilities; they model the complete behavior of the field.} In the next paragraphs I show how to construct weak observable currents from spacetime localized functions that can be calculated by integrating currents on a given codimension one embedded surface. 
Then I show that some of the constructed weak observable currents, in some cases all, can be promoted to observable currents. 
In the next section I briefly mention how a similar idea can be applied to bulk observables. 

Consider a spacetime localized measurement whose domain of sensitivity ${\cal S}$ is contained in a given codimension one surface $\Sigma$. Let me consider only functions of the following type
\[
\check{f}_\Sigma(\phi) \doteq 
\sum_\Sigma \; 
\tilde{\phi}^\ast \; \check{F} , \quad \mbox{ where } \tilde{\phi}^\ast \; 
\check{F}\neq 0  \mbox{ only in } 
{\cal S} \subset \Sigma.  
\]
Since the current in the integrand is not an observable current, $\check{f}_\Sigma$ may depend on the orientation of $\Sigma$ (which specifies the side used to evaluate it) in a way that may involve more than the sign. 
In addition, 
it may not be invariant under arbitrary deformations of the type $\Sigma \mapsto \Sigma' = \Sigma + \partial U'$. 
On the other hand, 
since the objective is to model physically relevant local properties, 
I do ask that when 
$\tilde{v}$ is in the null space of $\omega_{L, \Sigma}$ then
\[
\sum_\Sigma \; 
\tilde{\phi}^\ast \;
d\check{F} (\tilde{v} , \tilde{\phi}(\tau_\nu) ) = 0  . 
\]

Constructing an observable current that models a localized measurement has two primary motivations. 
The first one is that the outcome of a measurement 
with domain of sensitivity localized in the interior of the domain of interest ${\cal S}\subset \Sigma \subset U$ can be computed by integrating on a surface 
$\Sigma' = \Sigma + \partial U'$ contained in $\partial U$. A particular example
that highlights the importance of this property is to measure a property of the field at a particular spacetime region in terms of initial data. 
The second motivation is that, as shown in the previous subsection, 
there is a simple Poisson bracket that makes the space of observable currents a Poisson algebra. 
Thus, constructing observable currents corresponding to localized measurements
gives an algebraic structure to the space of these important physical observables. The relevance of this property for quantization is clear. In this respect, it should be noted that a quantization of this framework naturally leads to a spin foam model formulation of lattice field theory.

The basic strategy to construct an observable current $F \in {\rm wOC}_{\Sigma, U}$ 
that agrees with 
$\check{f}_\Sigma$ 
when integrated on $\Sigma$ is the following: \\
\begin{enumerate}
\item
Choose a reference solution $\phi_0 \in {\rm Sols}_U$. 
\item
Calculate the direct impact of measurement on the field. That is, find a vector field 
$\check{v}_\Sigma \in {\mathfrak X}(J^1 Y_\Delta |_\Sigma)$ solving the equation 
$\tilde{\phi}^\ast d\check{F} = - \tilde{\phi}^\ast \iota_{\check{v}_\Sigma} \Omega_L$. This vector field is unique up to the null directions of $\Omega_L|_{\tilde{\phi}(\Sigma)}$. 
Notice that this vector field induces a Hamiltonian vector field on the space of first-order data on $\Sigma$ (according to $\omega_{L,\Sigma}$). 
\item
Use the linearized field equations to calculate the indirect impact of the measurement caused by propagation. In other words, find a solution of the linearized field equations (\ref{linearizedFE}), 
$\tilde{v} \in {\mathfrak F}_{\tilde{\phi}(U)}$ that extends $\check{v}_\Sigma$. 
\item
The weak observable current $F_{\tilde{v}, k}$ defined in Subsection \ref{ConstrWocs}, with an appropriately adjusted integration constant will be the result of the construction. 
The chosen integration constant is 
$k_{\Sigma \check{F}}= \sum_\Sigma \; \tilde{\phi}_0^\ast \; \check{F}$. 
Since the objective is to define a weak observable current, the integration constant on other representatives of the homology class needs to be the same, and the integration constant for surfaces with the reversed orientation needs to have opposite sign. 
\end{enumerate}

{\em Recall that if the first variation $\tilde{v} \in {\mathfrak F}_{\tilde{\phi}(U)}$, resulting from the third step of the algorithm, is a commutator of first variations, then the weak observable current $F_{\tilde{v}, k}$ can be promoted an observable current as shown in Subsection \ref{ConstrWocs}.} 

Above, I gave a preliminary sketch of a subtle algorithm to construct weak observable currents corresponding to spacetime local measurement. In order for this vague idea to actually work several assumptions need to be verified. Now I mention the most prominent of them: 
At the core of this constructive procedure is solving the linearized field equations with data on $\Sigma$. Thus, $\Sigma$ must be of the appropriate type. 
$\Sigma$ may be a hypersurface appropriate for hosting initial data, or it may be appropriate for 
hosting a combination of initial conditions with spacelike boundary conditions that yield a solvable problem. 
If the domain of sensitivity of the measurement is ${\cal S}\subset \Sigma$, then one may require that $\check{v}_\Sigma$ vanishes outside ${\cal S}$. 
This requirement is appropriate if $\Sigma$ is a Cauchy surface for the linearized equations. For other types of hypersurfaces, the condition expressing that the measurement happened at ${\cal S}$ would be less transparent. 
As mentioned in Subsection \ref{ConstrWocs}, the domain of general $F_{\tilde{v}, K}$ may not be the whole $J^1Y_\Delta|_U$, and in general it is determined by $\tilde{v}|_\Sigma$. 
Since, in this construction $\tilde{v}$ was determined from the given function $\check{F}$ defined on the whole $J^1Y_\Delta|_\Sigma$, it follows that 
any restriction in the domain of definition of  
$F_{\tilde{v}, K_{\check{F}}}$ 
has origin purely on the behavior of solutions to the linearized field equations 
(\ref{linearizedFE}) 
with initial and/or boundary conditions set by the local measurement to be modeled (step two in the algorithm). 
Notice that these equations would be linear if a local section of interest is fixed, but observable currents need solutions of the equations at least in a neighborhood of the image of a solution of the field equations.

Examples of these observables can be exhibited analytically only for very simple cases. 
In reference \cite{meffGBFT} we show an algorithm within this formalism in a regular lattice for  solving the evolution problem of the scalar field. 
Since linearized scalar field theories are particular cases of the family of scalar field theories treated in  \cite{meffGBFT}, the mentioned algorithm solving the evolution problem may be applied 
to construct the vector fields required in the above construction of observable currents. The version of lattice gauge theory given in \cite{meffGBFT} can be placed on a regular lattice, and the theory may be linearized. Gauge invariance makes the system a bit more interesting, but the  evolution algorithm, with small modifications, is in essence equally straightforward. 

In numerical studies, this construction should not be difficult to implement.

In the previous subsection I introduced a Poisson bracket among observable currents. Here I 
consider the case of the bracket between two observable currents 
$F_{\tilde{v}, k_{\Sigma \check{F}}}$ and $G_{\tilde{w}, k_{\Sigma \check{G}}}$ modeling localized measurements with domains of sensitivity which are spacelike separated. 
I asume that the assumption translates into the domains of sensitivity of $\check{F}$ and $\check{G}$ 
fitting in the same Cauchy surface $\Sigma$. 
The result of the bracket is a new observable current 
$\{ F_{\tilde{v}, k_{\Sigma \check{F}}}, G_{\tilde{w}, k_{\Sigma \check{G}}}\}$
that when integrated on $\Sigma$ coincides with the bracket that one could have calculated using the initial data vector fields $\check{v}_\Sigma, \check{w}_\Sigma$ according to 
$\omega_{L, \Sigma}$. Thus, 
$\{ F_{\tilde{v}, k_{\Sigma \check{F}}}, G_{\tilde{w}, k_{\Sigma \check{G}}}\}=0$ if the domains of sensitivity of the localized measurements are spacelike separated. 

An abbreviated algorithm results from declaring that the second and third steps, 
constructing a Hamiltonian vector field associated with the localized measurement, 
are solved at once by an appropriate Green function acting on $\check{F}$. 
This connects to the idea of defining a Peierls bracket directly among $n-1$ cochains modeling localized measurement because for the bracket 
only the Hamiltonian vector fields are necessary. One should notice, however, that the result of such a bracket of two 
$n-1$ cochains with localized support 
would not be a 
chochain of the same type 
but an observable current. An detailed study of this subject is under preparation \cite{PeierlsDiscr}.

\section{Relation to bulk observables and the Peierls bracket}
\label{Bulk+Peierls}

In this section I briefly comment on modeling measurements that take place on $n$-dimensional regions. These physical observables, called bulk observables, are the primary focus of field theory. Consider a $n$-form $\check{F}^n$ with domain of sensitivity 
${\cal S}\subset U$. 
The Peierls bracket maps $\check{F}^n$ to the derivative operator 
$\{ \check{F} , \cdot \}_{\tiny\mbox{Peierls}}$. It is appropriate to assume that this derivative operator induces a vector field solving the linearized field equations $\tilde{v}_{\check{F}} \in {\mathfrak F}_{\tilde{\phi}(U)}$. 
This vector field can be used in the 
algorithm sketched in the previous subsection to construct a weak observable current from it and an 
integration constant $k_{\Sigma \check{F}}$. The result is a map from $n$-forms to weak observable currents 
\[
\check{F}^n \longmapsto F_{\tilde{v}_{\check{F}^n}, k_{\Sigma \check{F}}} . 
\]
Clearly, the Hamiltonian vector field of $F_{\tilde{v}_{\check{F}^n}, k_{\Sigma \check{F}}}$ 
is simply $\tilde{v}_{\check{F}^n}$. 
Thus, the map between bulk observables and weak observable currents (with their respective brackets) is an algebra homomorphism. 
A more detailed study of bulk observables and the Peierls bracket within this discrete formalism 
is in preparation \cite{PeierlsDiscr}. 

I recall two facts mentioned earlier that are relevant to the present subject: The resulting bracket between the two weak observable currents 
is an observable current. For some field theories all weak observable currents can be promoted to observable currents.

This construction shows that weak observable currents label equivalence classes of $n$-forms modeling bulk observables. The equivalence classes are defined by agreeing in their evaluation on solutions and having the same first-order effect on the field (inducing the derivative operator through the Peierls bracket). Thus, weak observable currents and observable currents appear in a central place in the algebraic structure of field theory and quantization. A first study of this subject in the continuum is under preparation \cite{OCsCont}. 

Again, I remind the reader that a quantization of this framework for classical field theory over a discretized spacetime 
leads to a spin foam model formulation of lattice field theory.

\section{Coarse OCs taken to a finer scale}
\label{CoarseGraining}

The discrete framework for multisymplectic field theory \cite{meffGBFT} at the basis for this proposal is equipped with coarse graining maps linking measuring scales related by refinement. 

If scale $\Delta'$ is finer than scale $\Delta$, to every history $\phi'$ at scale $\Delta'$ corresponds a history 
$\phi = \pi_{\Delta \Delta'}\phi'$ at scale $\Delta$. The coarse graining map $\pi_{\Delta \Delta'}$ acts by pull-back taking functions of coarse histories to finer scales. In this section I describe the relation between this coarse graining map and observable currents. 

The first thing to mention is that given 
$\phi_{\Delta'} \in {\rm Sols}_{U, \Delta'} \subset {\rm Hists}_{U, \Delta'}$ in general 
$\pi_{\Delta \Delta'}\phi_{\Delta'} \notin {\rm Sols}_{U, \Delta'} \subset {\rm Hists}_{U, \Delta'}$. 
Correcting the dynamics and the geometric structure 
at scale $\Delta$ to be compatible with the dynamics and structure at scale $\Delta'$ is possible. One way to do it is to compute a corrected action 
$S_{U, \Delta}(\phi_\Delta)$ as the extremum of $S_{U, \Delta'}$ on the set 
$\pi_{\Delta \Delta'}^{-1}(\phi_\Delta)$. Observables at scale $\Delta'$ 
can be constructed from the pull-back of functions of histories at scale $\Delta$, but the relevant domain is the space of solutions according to the corrected dynamics. 

The second thing that must be mentioned is that since coarse graining is inherently nonlocal, there is no coarse graining map $J^1Y_{\Delta'} \to J^1Y_{\Delta}$. Cochains on the discretization of the jet at scale $\Delta$ cannot be readily taken to finer scales. 

However, the mentioned difficulty may be solved. 
Coarse observable currents $F_\Delta$ can be taken 
to finer scales ; one way to do it is by the procedure stated below. 
\begin{enumerate}
\item 
Construct a function on histories using a codimension one surface, $f_{\Sigma , \Delta}$ and pull it back with the coarse graining map, $\pi_{\Delta \Delta'}^\ast f_{\Sigma , \Delta}$, 
obtaining a function of first-order data on $\Sigma$ at scale $\Delta'$. 
Notice that only if $F_\Delta$ is an observable current according to the $\Delta'$-corrected dynamics at scale $\Delta$, the obtained function of histories at scale $\Delta'$ would have the property of being insensitive to deformations of $\Sigma$ when evaluated at solutions of the $\Delta'$ dynamics. 
Additionally, only if $F_\Delta$ is an observable current according to the corrected dynamics, the resulting function of first-order data on $\Sigma$  will be independent of the null directions of $\omega_{L, \Sigma, \Delta'}$ (which is needed for the rest of the construction to work). 
\item 
Convert this function of first-order data into a Hamiltonian vector field 
$\check{v}_{\Sigma, \Delta'}$ 
on the space of first-order data at $\Sigma$ using $\omega_{L, \Sigma, \Delta'}$. 
\item
Complete the third step of the algorithm to construct observable currents modeling localized measurements. 
\item
Complete the fourth step of the mentioned algorithm. 
\end{enumerate}

An alternative way to coarse grain observable currents corresponding to localized measurement starts 
performing the first three steps in a single stroke, and then proceed to complete the last step. 
The first variation of the $\Delta'$ dynamics constructed in the first three steps, may be computed by means of 
the Peierls bracket briefly described at the end of Subsection \ref{LocalMeasurement}. 
The bracket corresponding to the action at scale $\Delta'$ acting on $\pi_{\Delta \Delta'}^\ast f_{\Sigma , \Delta}$ yields a 
locally Hamiltonian vector field according to the dynamics of scale $\Delta'$.

\section{Summary}
\label{Summary}



The concepts of observable current and weak observable current 
were introduced in a local framework for field theories over a discretized spacetime. 
Observable currents are local objects defined for any history. One family of examples of observable currents is composed by Noether currents. In this formalism they are written as $N_\xi= -\iota_{\tilde{v}_\xi} \Theta_L$, 
where the vector field $\tilde{v}_\xi$ is the generator in the jet of the symmetry 
indicated by the symmetry group generator $\xi\in Lie({\cal G})$.  
A much larger family of examples has elements of the type $F=\iota_{\tilde{w}} \iota_{\tilde{v}} \Omega_L$, where the vector fields are first variations ${\tilde{v}}, {\tilde{w}}\in {\mathfrak F}_{\tilde{\phi}(U)}$. 

Weak observable currents modeling localized measurements of the field were implicitly constructed, and they were shown to separate distinct neighboring solutions. 
Moreover, it was shown that Hamiltonian weak observable currents were sufficient to distinguish physically distinct solutions.  

Some weak observable currents constructed to model localized measurement can be 
``improved'' to become local 
and defined for every history instead of being defined only for solutions. 
This is the case when 
their defining locally Hamiltonian vector field is a commutator (or a linear combination of commutators). 
If in a field theory all locally Hamiltonian vector fields are generated by commutators 
(of locally Hamiltonian vector fields), then 
every localized measurement can be modeled by means of an observable current. 
Even if not all localized measurements can be modeled using observable currents, the separability condition 
(\ref{dOC}) may be satisfied using only observable currents. 
This is the case for a field theory such that 
every first variation that is symplectically orthogonal to the space generated by commutators (of locally Hamiltonian vector fields) is necessarily zero. 
A study towards characterizing the field theories that have this property is underway. 

In Subsections \ref{ConstrWocs}, \ref{LocalMeasurement} I mentioned that the domain of definition of observable currents may not be the whole discrete jet bundle $J^1Y_\Delta|_U$. 
In particular, if a weak observable current modeling a localized measurement is constructed following the algorithm described in those sections, a primary ingredient is a solution of the linearized field equations. Thus, the extension of the domain of definition of these weak observable currents is dictated by those equations and the initial and/or boundary conditions. 
In the case that one of the mentioned weak observable currents is promoted to an observable current, 
since the replacing observable current is constructed in terms of locally Hamiltonian vector fields related to the first one by a commutator, 
the domain of definition of the observable current has to be contained in the domain of definition of the original weak observable current. 
For other strong motivations to consider observables defined on subdomains of the space of the field configurations, see Khavkine's work on generalized local observables \cite{Khavkine}.  

More than explicit construction of observable currents, what is needed is: 
(i) 
To have a study of existence and uniqueness for weak observable currents modeling different types of measurements. 
(ii) 
In the case of localized measurements, one should find physically reasonable conditions on the seed cochain $\check{F}$ (or $\check{F}^n$) that guarantee existence and uniqueness of weak observable currents modeling the measurement. 
It would be interesting to investigate 
possible relations between the study of existence of weak observable currents in local domains of definition and the related issue investigated in \cite{DKN} in a more stringent context. 
(iii)
To have a characterization of which type of seed cochain $\check{F}$ (or $\check{F}^n$) describe 
localized measurements that can be modeled by observable currents. 
(iiii) 
To have a study of which theories are such that observable currents distinguish physically distinct solutions. 
In this moment, it is known that this happens for field theories in which  
every first variation that is presymplecticaly orthogonal to the space generated by commutators 
of locally Hamiltonian first variations 
is necessarily in the null space of the presymplectic form. It would be desirable to know if this happens for field theories of interest; it would be great to be able to characterize Lagrangians leading to field theories with this property.

In continuum treatments of multisymplectic field theory 
\cite{MSobs, DynObsEtc, Kanatchikov} 
there are proposals related to the notion of observable current introduced in this article, but there are important differences. 
Also, they have developed aspects that have not been treated here, and some of the results given in this article do not have an analog in the continuum. 
The local (multisymplectic) equation relating observable currents to Hamiltonian vector fields presented in the work $dF=-\iota_{\tilde{v}} \Omega_L + \alpha_{\tiny \Sigma}$, 
can be taken to the continuum. There,  it 
extends notions similar to that of Hamiltonian observable current that had been previously investigated. 
The term $\alpha_{\tiny \Sigma}$ is innocuous in 
the sense that it does not affect the mentioned induced relation in the space of solutions, 
but its presence allows for a large class of Hamiltonian 
observable currents that would not be so without it. 
Direct analogs of the notions of observable current, weak observable current and the Poisson algebra introduced in this article can be written in the continuum; an account of this study is in preparation \cite{OCsCont}. 
Localized measurement was also included in this presentation, and I mentioned that the associated algebraic structure could be constructed directly by means of a Peierls bracket. This work is still in progress \cite{PeierlsDiscr}, and it would be interesting to compare its results with recent work of Forger \cite{Forger}.

In Section \ref{Bulk+Peierls}, I mentioned that observable currents labeled the equivalence classes of bulk observables that naturally appear in the algebraic structure of classical field theories and emphasized the importance of this structure in quantization. This avenue is under study 
\cite{OCsCont}. 
It would be interesting to study the relation of the resulting framework with Costello's recent work on factorization algebras in field theory \cite{Costello}.

Since multisymplectic geometry is the higher dimensional counterpart of symplectic geometry, the associated algebraic structures have been of interest in the study of 
higher category theory. In particular, an algebra of currents resulting from the multisymplectic framework has been explored already from that point of view \cite{Ncat}. The different notions of observable current developed here could be interesting also in that context.

\section*{Acknowledgements} 
This work was partially supported by grant PAPIIT-UNAM IN109415. 
The definition of the product for observable currents leading to a Poisson algebra was found during a discussion with Alberto Molgado and Jasel Berra. 
I am grateful to 
Igor Khavkine, Claudio Meneses, Robert Oeckl, Michael Reisenberger, Juan Daniel Reyes and 
José A. Vallejo for discussions during critical stages of the project.


\end{document}